\documentclass[aps,prl,twocolumn,showpacs,amsmath,amsmath,amssymb]{revtex4-1}

\usepackage{amsmath,amssymb,graphicx,bm,psfrag,bbold,float,color,datetime,appendix}

\usepackage{color}

\usepackage[mathscr]{eucal}

\DeclareMathAlphabet{\mathpzc}{OT1}{pzc}{m}{it}

\begin{document}

\title{Theoretical Description of Micromaser in the Ultrastrong-Coupling Regime}

\author{Deshui Yu$^{1}$, Leong Chuan Kwek$^{1,2,3,4}$, Luigi Amico$^{1,5,6}$, and Rainer Dumke$^{1,7,\ast}$}

\affiliation{$^{1}$Centre for Quantum Technologies, National University of Singapore, 3 Science Drive 2, Singapore 117543, Singapore}

\affiliation{$^{2}$Institute of Advanced Studies, Nanyang Technological University, 60 Nanyang View, Singapore 639673, Singapore}

\affiliation{$^{3}$National Institute of Education, Nanyang Technological University, 1 Nanyang Walk, Singapore 637616, Singapore}

\affiliation{$^{4}$MajuLab, CNRS-UNS-NUS-NTU International Joint Research Unit, UMI 3654, Singapore}

\affiliation{\mbox{$^{5}$CNR-MATIS-IMM \& Dipartimento di Fisica e Astronomia, Universit\'a Catania, Via S. Soa 64, 95127 Catania, Italy}} 

\affiliation{$^{6}$INFN Laboratori Nazionali del Sud, Via Santa Sofia 62, I-95123 Catania, Italy}

\affiliation{$^{7}$Division of Physics and Applied Physics, Nanyang Technological University, 21 Nanyang Link, Singapore 637371, Singapore}

\date{\today}

\begin{abstract}
We theoretically investigate an ultrastrongly-coupled micromaser based on Rydberg atoms interacting with a superconducting LC resonator, where the common rotating-wave approximation and slowly-varying-envelope approximation are no longer applicable. The effect of counter-rotating terms on the masing dynamics is studied in detail. We find that the intraresonator electric energy declines and the microwave oscillation frequency shifts significantly in the regime of ultrastrong coupling. Additionally, the micromaser phase fluctuation is suppressed, resulting in a reduced spectral linewidth.
\end{abstract}


\maketitle

\textit{Introduction.} The recent dramatic progress in superconducting quantum circuits enables the fundamental study of ultrastrong interfaces between particles and electromagnetic fields. Recently, artificial atoms have been linked to the superconducting microwave resonators in the ultrastrong coupling regime~\cite{PRA:Ciuti2005,PRA:Ciuti2006,AnnPhys:Devoret2007,NatPhys:Niemczyk2010}. Furthermore, there is large experimental effort to store neutral atoms in the vicinity of superconducting circuits~\cite{EPL:Haroch2008,PRL:Shimizu2009,PRA:Siercke2012,PRA:Patton2013,NatComm:Bernon2013,PRA:Yu2016}, paving the way towards masing/lasing physics in the regime of ultrastrong coupling. Conventional masers and their optical counterpart, lasers, rely on the weak/strong particle-resonator interactions, for which the well-known rotating-wave and slowly-varying-envelope approximations are usually employed to explore the system characteristics~\cite{PR:Lamb1964,PRL:Scully1966,PR:Stenholm1969,PRA:Benkert1990,PRA:Kolobov1993,RMP:Davidovich1996,PRA:Filipowicz1986,PRL:Rempe1987,PRL:An1994,Nature:McKeever2003,JOptSocAmB:Clemens2004,PRA:FangYen2006,JOSAb:Yu2016}.

The rotating-wave approximation fails when the coupling strength between electromagnetic fields and active particles is close to the electromagnetic frequency~\cite{PhysScr:Larson2007,PhysicaA:Ford1997}. Actually, the counter-rotating terms take apparent effects in some circumstances, for examples, quantum chaos~\cite{JETP:Belobrov1976,PRL:Milonni1983,PRA:Peng1992,PRL:Emary2003,PRE:Emary2003}, Landau-Zener transitions of a qubit in circuit QED~\cite{EurophysLett:Saito2006,PRB:Saito2007}, entanglement of spin anticorrelated state~\cite{PRA:Jing2009,PhysScr:Ficek2010}, maser/laser dynamics~\cite{PRA:Vyas1986,OptCommun:Zela1997,JModOpt:Liu1998,Carter:Zela2001}, \textit{et al.}, even when the atom-field interaction is much smaller than the electromagnetic frequency. Also, the slowly-varying-envelope approximation becomes invalid because such an ultrastrong atom-field interaction significantly disturbs the electromagnetic field at a rate comparable to the carrier frequency. Owing to the large cavity-mode volume and small atomic dipole moments, the ultrastrong-coupling regime has never been experimentally accessed before in the conventional maser/laser physics.

In this paper, we study a hybrid micromaser system based on Rydberg atoms interacting with a superconducting LC resonator. The atom-resonator coupling strength is tuned from the weak- to the ultrastrong-coupling regime, where the ordinary rotating-wave and slowly-varying-envelope approximations are inapplicable. The resulting micromaser properties in different regimes exhibit distinguishing characteristics. The nonconventional behavior in the regime of ultrastrong coupling is attributed to the counter-rotating terms.

\textit{Physical model.} We consider a micromaser system, where $^{87}$Rb Rydberg atoms interact with a single-mode superconducting electromagnetic cavity (see Fig. 1a). The resonator, operating at mK temperatures to suppress the thermal background, is designed for an experimentally accessible resonance frequency of $\omega_{0}=2\pi\times10.0$ GHz. For a LC resonator, as an example, this can be achieved by using an inductor of $L=0.7$ $\mu$H~\cite{arXiv:Sarabi2017} and a capacitor $C$ composed of a pair of parallel cylinders with a radius of $a=4.0$ $\mu$m, length of $\Delta l=4.1$ $\mu$m along the $z$-axis, and interaxial distance of $d=8.4$ $\mu$m in the $x$-direction, resulting in a capacitance of $C=362.0$ aF. The smallest intercylinder distance is $d=b-2a=0.4$ $\mu$m ($d\ll2a$). The atoms propagate along the $x$-direction in vicinity to the capacitor plates and couple to the $z$-direction fringe field $\eta{\cal{E}}$. The parameter $\eta$ measures the ratio of fringe field to central field ${\cal{E}}$ inside capacitor. Changing the distance between atoms and capacitor varies $\eta$ (see Fig. 1b) and hence tunes the atom-resonator coupling strength.

\begin{figure*}
\includegraphics[width=17.5cm]{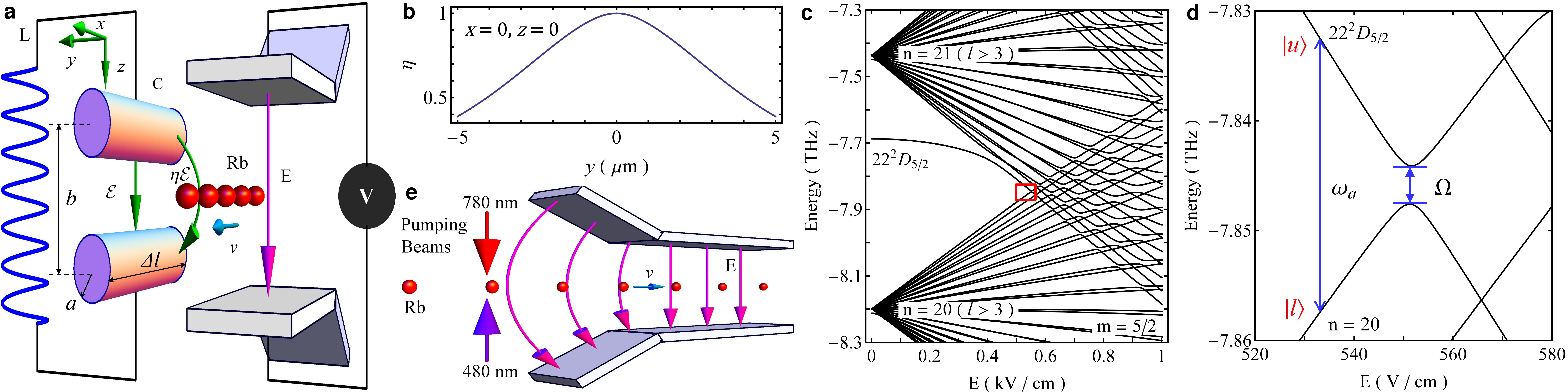}\\
\caption{(Color online) Micromaser scheme. (a) Slowly moving Rydberg atoms interact with a superconducting LC resonator via the fringe field $\eta{\cal{E}}$ of capacitor. A static electric field $E$ is applied to tune the atomic levels. (b) $\eta$ vs. the distance $y$ with $x=z=0$. The origin ($x=y=z=0$) of coordinates is set at the central point of the cylindrical capacitor $C$. (c) dc Stark map of $^{87}$Rb. The detail inside the rectangle frame is shown in (d). (e) Schematic of pumping process. Before entering the capacitor, the atoms are excited onto the Rydberg $22^{2}D_{5/2}(m=\frac{5}{2})$ state via the two-photon transition.}\label{Fig1}
\end{figure*}

Following~\cite{PRA:Blais2004}, the quality factor of superconducting LC resonator is estimated as $Q=\frac{\omega_{0}}{\kappa}=2\times10^{4}$, where $\kappa=2\pi\times0.5$ MHz denotes the rate of energy loss within resonator. According to Kirchoff's Laws, the wave equation for the electric field ${\cal{E}}$ inside capacitor is derived as
\begin{equation}\label{ElectricFieldEq}
\frac{d^{2}}{dt^{2}}{\cal{E}}(t)+\kappa\frac{d}{dt}{\cal{E}}(t)+\omega^{2}_{0}{\cal{E}}(t)+\frac{1}{\epsilon_{0}}\frac{d^{2}}{dt^{2}}{\cal{P}}(t)=0,
\end{equation}
where ${\cal{P}}$ is the polarization density of gain medium and $\epsilon_{0}$ is the vacuum permittivity. The above second-order differential equation describes also common maser/laser systems~\cite{Book:Scully}. The medium polarization ${\cal{P}}$ governs the dispersion and absorption/amplification of microwave. Unlike the traditional maser/laser theory, here we do not utilize the slowly-varying-envelope approximation to convert Eq.~(\ref{ElectricFieldEq}) into a first-order form because the ultrastrong atom-resonator coupling varies the amplitude of electromagnetic field rapidly, comparable with the carrier frequency.

An external electrostatic field $E$ along the $z$-direction is applied on the atoms to adjust the energy spectrum of $^{87}$Rb (see Fig. 1a). Following the condition given by the resonator, a suitable candidate for the maser transition is shown in the dc stark map (see Fig. 1c and 1d) of $^{87}$Rb around the $22^{2}D_{5/2}(m=\frac{5}{2})$ Rydberg state, calculated according to~\cite{PRA:Zimmerman1979,PRA:Marinescu1994}. It is seen that at $E=550.7$ V/cm there exists an avoided-level crossing occurring between two adiabatic curves starting with $22^{2}D_{5/2}(m=\frac{5}{2})$ and a manifold $\Phi_{n=20}$ composed of a set of $|n=20,l\geq3,j=l\pm\frac{1}{2},m=\frac{5}{2}\rangle$ ($n$, $l$, $j$, and $m$ are the principal, orbital, total angular momentum, and magnetic quantum numbers, respectively) at $E=0$. The energy gap of anticrossing is given by $\Omega=2\pi\times3.2$ GHz. Two eigenstates $|u\rangle$ and $|l\rangle $, far enough away from the energy-level anticrossing, in different adiabatic curves are chosen to form the maser transition with a frequency of $\omega_{a}$. The large energy separations between $|u\rangle$, $|l\rangle$ and any other adiabatic eigenstates strongly suppress the latter's effect on the masing dynamics. Thus, the Rydberg atom can be simplified as a two-level system consisting of $|u\rangle$ and $|l\rangle$.

The coherent atom-resonator interface is governed by the Hamiltonian~\cite{PRA:Yu2016-2}
\begin{equation}\label{HamiltonianH}
H=\hbar\sum_{j}\theta(t-t_{j})\left[\frac{\omega_{a}}{2}\sigma^{(j)}_{z}+\frac{\Omega_{\textrm{eff}}(t)}{2}\sigma^{(j)}_{x}\right],
\end{equation}
where the unit step function $\theta(t-t_{j})$ denotes the $j$-th atom starts interacting with the superconducting LC resonator at the time $t_{j}$. The maser-transition frequency $\omega_{a}$ can be tuned by adiabatically varying the electrostatic field $E$. The time-dependent effective driving frequency is defined as
\begin{equation}\label{ElectricFieldEq}
\Omega_{\textrm{eff}}(t)=\Omega-2\eta\mathpzc{d}_{0}{\cal{E}}(t)/\hbar.
\end{equation}
$\mathpzc{d}_{0}=|\langle u|\mu|l\rangle|=385$ $ea_{0}$~\cite{PRA:Yu2016-2} gives the atomic dipole moment of the masing transition, where $\mu$ is the electric dipole operator of an atom. The Pauli matrix operators for the $j$th atom are defined as $\sigma^{(j)}_{x}=\sigma^{(j)\dag}_{-}+\sigma^{(j)}_{-}$ and $\sigma^{(j)}_{z}=\sigma^{(j)}_{uu}-\sigma^{(j)}_{ll}$ with $\sigma^{(j)}_{uu}=(|u\rangle\langle u|)_{j}$, $\sigma^{(j)}_{ll}=|l\rangle\langle l|$, and $\sigma^{(j)}_{-}=(|l\rangle\langle u|)_{j}$.

\begin{figure*}
\includegraphics[width=17.5cm]{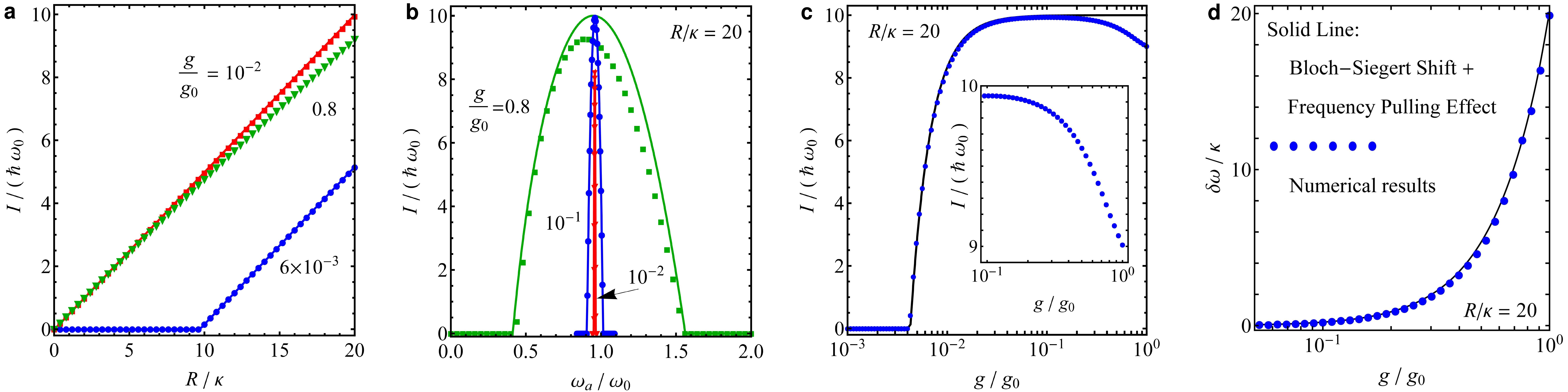}\\
\caption{(Color online) (a) Time-averaged energy $I$ (in units of $\hbar\omega_{0}$) of the steady-state electric field ${\cal{E}}^{s}(t)$ changing with the pumping rate $R$ for different coupling strengths $g$. The atomic-transition frequency $\omega_{a}$ is shifted to $\sqrt{\omega^{2}_{0}-\Omega^{2}}$. (b) The energy $I$ is depicted in dependence on $\omega_{a}$ for different $g$ with $R/\kappa=20$. (c) $I$ vs. $g$ with $R/\kappa=20$ and $\omega'_{a}=\omega_{0}$. For comparison, we also insert the results of conventional maser/laser theory (solid lines) in (b) and (c). The calculated frequency difference $\delta\omega=\omega_{c}-\omega_{0}$ between the microwave frequency $\omega_{c}$ and the resonator frequency $\omega_{0}$ is displayed in (d), where the solid line indicates the expected frequency shift taking into account the Bloch-Siegert shift and frequency pulling effect. The other parameters in (d) are same to (c).}\label{Fig2}
\end{figure*}

The pumping scheme is illustrated in Fig. 1e. The electrostatic field $E$ is produced by another constant-voltage-biased capacitor, whose inner region is divided into a wedge-shaped and a parallel-plate part~\cite{PRL:Vliegen2004}. The connection between plates of different parts should be arch-shaped to suppress the edge effect. The static electric field gradually increases from zero to $E$ in the wedge-shaped region and remains at $E$ in the parallel-plate region. Atoms in the ground state $5^{2}S_{1/2}(m=\frac{1}{2})$ fly into the capacitor individually. We assume that the atoms have the same velocity $v$ and are equally separated with a distance of about $\Delta l$ (large enough to suppress the effect of interatomic interaction on the masing dynamics), leading to only one atom $N=1$ interacting with the resonator at a time. The realization of such a deterministic source of atoms increases the complexity of experimental implementation. Before entering the electrostatic-field capacitor, the atoms are excited into $22^{2}D_{5/2}(m=\frac{5}{2})$ via the light fields at 475 nm and 780 nm. Then, they adiabatically transit to the upper masing level $|u\rangle$, resulting in the population inversion in the parallel-plate region. Since we do not focus on the intensity fluctuation of maser field in this work, a Poissonian-pumping statistic is assumed for simplicity (see Appendix A). However, it should be noted that a sub-Poissonian or regular pumping can lead to the memory effects and reduce the low-frequency intensity fluctuation of the maser/laser field~\cite{PRA:Benkert1990,PRA:Kolobov1993,RMP:Davidovich1996}.

The injection values of atomic variables are derived at 
$\bar{\sigma}_{z}=\langle\sigma^{(j)}_{z}\left(t=t_{j}\right)\rangle=(\Omega^{2}-\Omega'^{2})/(\Omega'^{2}+\Omega^{2})$,
$\bar{\sigma}_{x}=\langle\sigma^{(j)}_{x}\left(t=t_{j}\right)\rangle=2\Omega'\Omega/(\Omega'^{2}+\Omega^{2})$, and $\bar{\sigma}_{y}=\langle\sigma^{(j)}_{y}\left(t=t_{j}\right)\rangle=0$, where the $y$-component Pauli matrix operator $\sigma^{(j)}_{y}=-i(\sigma^{(j)\dag}_{-}-\sigma^{(j)}_{-})$ and we have defined $\omega'_{a}=\sqrt{\omega^{2}_{a}+\Omega^{2}}$ and $\Omega'=\omega'_{a}-\omega_{a}$. The zero-Kelvin lifetimes of $|u\rangle$ and $|l\rangle $ are of the order of 10 $\mu$s~\cite{JPB:Branden}. However, the Rydberg-state lifetimes are strongly reduced when the atom is nearby the chip surface due to the stray electric fields~\cite{PRA:HermannAvigliano2014} and ionization~\cite{EPJST:Kohlhoff2016}. According to~\cite{PRA:Crosse2010,JPB:Fabre1983}, the decay rates of $|u\rangle$ and $|l\rangle $ are reasonably chosen to be $\Gamma=2\pi\times30.0$ MHz for the typical atom-surface distance of the order of 1 $\mu$m (corresponding to $0.1\leq\eta<1$) in this work (see Fig. 1b). In addition, the Casimir-Polder shifts hardly affect the atomic spectrum.

Applying the Heisenberg-Langevin approach~\cite{PRA:Yu2010a,PRA:Yu2010b,PRA:Yu2011}, we obtain the equations of motion for the $c$-number atomic variables from $H$ (see Appendix A)
\begin{subequations}\label{rhoEq}
\begin{eqnarray}
\frac{d}{dt}{\cal{U}}(t)&=&R\bar{\sigma}_{x}-\Gamma{\cal{U}}(t)-\omega_{a}{\cal{V}}(t)+{\cal{F}}_{{\cal{U}}}(t),\\
\frac{d}{dt}{\cal{V}}(t)&=&-\Gamma{\cal{V}}(t)+\omega_{a}{\cal{U}}(t)-\Omega_{\textrm{eff}}(t){\cal{W}}(t)+{\cal{F}}_{{\cal{V}}}(t),~~~\\
\frac{d}{dt}{\cal{W}}(t)&=&R\bar{\sigma}_{z}-\Gamma{\cal{W}}(t)+\Omega_{\textrm{eff}}(t){\cal{V}}(t)+{\cal{F}}_{{\cal{W}}}(t),
\end{eqnarray}
\end{subequations}
where ${\cal{U}}(t)$, ${\cal{V}}(t)$, and ${\cal{W}}(t)$ are the $c$-number counterparts of macroscopic atomic operators
$U(t)=\sum_{j}\theta_{j}(t-t_{j})\sigma^{(j)}_{x}$, $V(t)=\sum_{j}\theta_{j}(t-t_{j})\sigma^{(j)}_{y}$, and $W(t)=\sum_{j}\theta_{j}(t-t_{j})\sigma^{(j)}_{z}$, respectively. $R=Nr_{a}$ is the injection rate of atoms with $r_{a}=v/\Delta l$. For the value of $R$ interested in this work, the atomic velocity $v$ is of the order of $10^{2}$. The corresponding atom-resonator interaction time is tens of ns, longer than the decay time $\Gamma^{-1}$ of masing states, avoiding the transit effect from the finite atom-resonator interaction time~\cite{PRA:Yu2008}. ${\cal{F}}_{{\cal{U}},{\cal{V}},{\cal{W}}}(t)$ are the Langevin noises for different atomic variables. The atomic polarizability in Eq.~(\ref{ElectricFieldEq}) is given by ${\cal{P}}(t)=\eta\mathpzc{d}_{0}{\cal{U}}(t)/V_{\textrm{eff}}$ with the effective volume $V_{\textrm{eff}}=Sd$ and $S=Cd/\epsilon_{0}$.

Neglecting the Langevin forces ${\cal{F}}_{{\cal{U}},{\cal{V}},{\cal{W}}}(t)$, Eqs.~(\ref{ElectricFieldEq}) and (\ref{rhoEq}) describe the semiclassical dynamics of microwave laser without rotating-wave and slowly-varying-envelope approximations. Due to the counter-rotating terms, the hybrid system does not reach a stationary state and the analytical investigation is unpractical. Here we employ the well-known fourth-order Runge-Kutta method to numerically solve the masing dynamics. For an arbitrary initial state, the dissipative atom-resonator system arrives at an equilibrium state after a long time $t_{0}\gg\textrm{max}(\Gamma^{-1},\kappa^{-1})$. We use ${\cal{E}}^{s}(t)$, ${\cal{U}}^{s}(t)$, ${\cal{V}}^{s}(t)$, ${\cal{W}}^{s}(t)$, and $\Omega^{s}_{\textrm{eff}}(t)$ to denote the steady-state solutions of electric field, atomic variables, and effective Rabi frequency, respectively.

According to the circuit quantum electrodynamics~\cite{PRA:Yu2016-2} and full quantum theory of traditional masers/lasers~\cite{PRA:Benkert1990,PRA:Kolobov1993,RMP:Davidovich1996}, the magnitude of quantized electric field (produced by the charge of $Q=\sqrt{2C\hbar\omega_{0}}$) inside the resonator is given by ${\cal{E}}_{\textrm{vac}}=\sqrt{2\hbar\omega_{0}/(Cd^{2})}$, where we have used the fact that $d\ll2a,\Delta l$. We choose the parameter $g=\frac{\eta\mathpzc{d}_{0}{\cal{E}}_{\textrm{vac}}}{2\hbar}=\eta\mathpzc{d}_{0}\sqrt{\frac{\omega_{0}}{2\hbar\epsilon_{0}V_{\textrm{eff}}}}$, which is proportional to $\eta$, to measure the atom-resonator coupling strength. For $\eta=1$, $g$ reaches the maximum given by $g_{0}=2\pi\times1.2$ GHz. We define the cooperativity parameter of the hybrid system ${\cal{C}}=2g^{2}/(\Gamma\kappa)=\eta^{2}{\cal{C}}_{0}$, with the maximum ${\cal{C}}_{0}=2.0\times10^{5}$. ${\cal{C}}^{-1}$ gives the critical atom number for the masing/lasing output in the weak- or strong-coupling regime~\cite{PhysScr:Kimble1998}.

\textit{Electric-field Energy.} For the hybrid system in the steady state, the time-averaged energy $I$ of electric field confined in LC resonator is given by $I=\epsilon_{0}V\langle\left[{\cal{E}}^{s}(t)\right]^{2}\rangle_{t}=\lim_{T\to\infty}\frac{\epsilon_{0}V}{T}\int_{t_{0}}^{t_{0}+T}[{\cal{E}}^{s}(t)]^{2}dt$. In the conventional maser/laser theory~\cite{Book:Ohtsubo}, $I$ can be analytically derived as $I=I_{s}(R/R_{th}-1)$, where the saturation energy is $I_{s}=\hbar\omega_{0}\frac{\Gamma^{2}}{4g^{2}}\frac{\omega'^{2}_{a}}{\omega^{2}_{a}}\left[1+\left(\frac{\omega'^{2}_{a}-w_{0}}{\kappa/2+\Gamma}\right)^{2}\right]$, and the pumping threshold is $R_{th}=\frac{\kappa\Gamma^{2}}{2g^{2}}\frac{\omega'^{2}_{a}}{\omega^{2}_{a}}\left[1+\left(\frac{\omega'^{2}_{a}-w_{0}}{\kappa/2+\Gamma}\right)^{2}\right]$. However, as we will see below, this result is only valid in the weak- and strong-coupling regimes where the common rotating wave approximation is applicable.

Figure 2a displays $I$ as a function of the pumping rate $R$ for several different coupling strengths $g$. A masing threshold is present in the weak-coupling limit (${\cal{C}}<1$). In contrast, the masing dynamics exhibits thresholdless behavior in the strong-coupling regime (${\cal{C}}\gg1$ and $g/g_{0}\ll1$) and $I$ raises linearly as $R$ is increased from zero. Our numerical results agree with that of the traditional maser/laser theory in both weak- and strong-coupling regimes.

When $g$ is further enhanced to the regime of ultrastrong coupling ($g/g_{0}\sim1$), $I$ becomes lower than that of the hybrid system operating in the strong-coupling regime for increasing $R$. This degradation coincides with the prediction in~\cite{PRA:Vyas1986}, i.e., the counter-rotating terms reduce the maser/laser gain. In addition, $I$ is no long linearly proportional to $R$.

\begin{figure*}
\includegraphics[width=17.5cm]{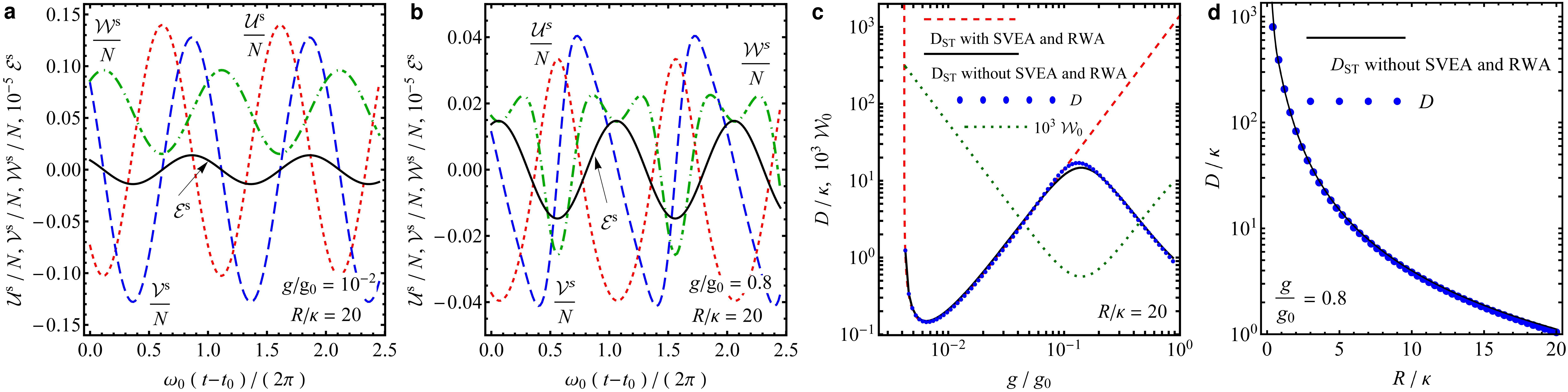}\\
\caption{(Color online) (a) and (b): Steady-state oscillations of the internal electric field ${\cal{E}}^{s}$ and the atomic variables ${\cal{U}}^{s}$, ${\cal{V}}^{s}$, and ${\cal{W}}^{s}$ in the strong- and ultrastrong-coupling regimes. (c) Spectral linewidth $D$ as a function of $g$ with $R/\kappa=20$. (d) Dependence of $D$ on the pumping rate $R$ with $g/g_{0}=0.8$. For comparison, we also insert the Schawlow-Townes limit $D_{ST}$.}\label{Fig3}
\end{figure*}

Figure 2b illustrates the energy $I$ changing with the atomic-transition frequency $\omega_{a}$. In the limit of $\omega_{0}\gg g$, $I$ maximizes at $\omega'_{a}=\omega_{0}$, i.e., the dc-Stark shifted atomic transition is resonant to the LC resonator. As $g$ is increased, the masing region is broadened symmetrically. For a very large atom-resonator coupling, the microwave oscillation can be maintained even without the overlap between the resonator-mode spectrum and the gain spectrum, $|\omega'_{a}-\omega_{0}|>(\kappa+\Gamma)$. In the regime of ultrastrong coupling, the dependence of $I$ on $\omega_{a}$ around $\omega_{0}$ is no longer symmetric, owing to the avoided energy-level crossing between $|u\rangle$ and $|l\rangle$ when $\omega_{a}$ approaches $\Omega$ (see Fig. 1d). Our numerical result violates the conventional maser/laser theory because of the counter-rotating terms. The maximum of $I$ is reduced and shifted to the low-frequency side of $\omega_{0}$.

We further consider the dependence of $I$ on $g$ with the resonant atom-resonator interaction, i.e., $\omega'_{a}=\omega_{0}$, for a fixed $R$ (see Fig. 2c). The conventional maser/laser theory predicts that $I$ is saturated when ${\cal{C}}\gg1$. However, our numerical results shows clearly that $I$ starts to decline from $g/g_{0}\approx0.1$, i.e., $g\sim10^{-2}\omega_{0}$, indicating $g\ll\omega_{0}$ is not a sufficient condition for the validity of common rotating-wave approximation. Actually, the decline of $I$ in the regime of ultrastrong coupling is caused by the increasing effective population inversion. Energy and atom number conservations give
\begin{equation}\label{relation1}
R\bar{\sigma}_{z}-\Gamma{\cal{W}}_{0}=2\kappa I/(\hbar\omega_{a}),
\end{equation}
where ${\cal{W}}_{0}\equiv\langle{\cal{W}}^{s}(t)\rangle_{t}$ is the averaged population inversion. When $g/g_{0}$ approaches 1, the counter-rotating terms disturb the conventional rotating atom-resonator interaction, resulting in the rise of ${\cal{W}}_{0}$ (see below) and the decrease of $I$ for a fixed pumping rate.

One can also derive the masing oscillation frequency $\omega_{c}$ from the numerical simulation. As shown in Fig. 2d, although the superconducting cavity is resonantly coupled to the dc-Stark shifted atomic transition when $\omega_{0}=\omega'_{a}$, $\omega_{c}$ is significantly shifted from the characteristic frequency $\omega_{0}$ of the LC resonator in the ultrastrong-coupling regime. It has been pointed out that the counter-rotating terms can give rise to a shift in the true resonance frequency of the atomic transition, i.e., Bloch-Siegert shift~\cite{PR:Bloch1940,PRB:Shirley1965,PRL:FornDiaz2010,PRL:Tuorila2010} $\Delta\omega_{BS}=(\eta\mathpzc{d}_{0}{\cal{E}}_{0})^{2}/(4\hbar^{2}\omega_{a})$, where ${\cal{E}}_{0}$ is the micromaser amplitude. This energy-level shift, which is much larger than the cold-cavity linewidth $\kappa$ and the atomic-transition linewidth $\Gamma$, additionally induces the cavity frequency pulling effect~\cite{Book:Lasers-Siegman} and the resulting oscillation frequency is given by $\omega_{c}\approx\omega_{0}+\Delta\omega_{BS}/(1+2\Gamma/\kappa)$. Such an apparent mismatch between $\omega_{c}$ and $\omega_{0}=\omega'_{a}$ is the other characteristic of ultrastrong atom-resonator coupling.

\textit{Spectral Linewidth.} For the system in the steady state, the atoms periodically exchange energy with the superconducting resonator. As shown in Figs. 3a and 3b, the steady-state electric field ${\cal{E}}^{s}(t)$ primarily exhibits the sinusoidal behavior in different atom-resonator coupling regimes ${\cal{E}}^{s}(t)={\cal{E}}_{0}\cos\omega_{c}t$. In contrast, the sinusoidal oscillations of macroscopic atomic variables are significantly distorted in the ultrastrong-coupling regime because of the enhanced counter-rotating terms. Nevertheless, all different variables oscillate at the same frequency $\omega_{c}$.
We should note that although the steady-state micromaser mainly presents a sinusoidal oscillation, it does not indicate the applicability of slowly varying envelope approximation in the ultrastrong-coupling regime.

To investigate the spectral linewidth of the Rydberg micromaser, we express all variables as a sum of the steady-state part and a small fluctuating term, for example, ${\cal{E}}(t)={\cal{E}}^{s}(t)+\delta{{\cal{E}}}(t)$. By Fourier transform, e.g., $\delta{\cal{E}}(\omega)=\int_{-\infty}^{\infty}\delta{\cal{E}}(t)e^{-i\omega t}dt$, Eqs.~(\ref{rhoEq}) are reduced to a set of algebraic equations for the fluctuations with the Langevin forces ${\cal{F}}_{{\cal{U}},{\cal{V}},{\cal{W}}}(\omega)$ as the noise sources. The linewidth of internal microwave is attributed to the phase fluctuation
\begin{equation}\label{phasefluctuation}
\delta\varphi(\omega)=\frac{{\cal{E}}(\omega)-{\cal{E}}^{\ast}(-\omega)}{2i{\cal{E}}_{0}},
\end{equation}
for $\omega$ around $\omega_{c}$. $\delta\varphi(\omega)$ is $\delta$ correlated in frequency, i.e.,
\begin{equation}\label{deltacorrelated}
\langle\delta\varphi(\omega)\delta\varphi(\omega')\rangle=\frac{D}{(\omega-\omega_{c})^{2}}\delta(\omega+\omega'),
\end{equation}
because of the $\delta$-correlated feature of Langevin forces and the spectral linewidth $D$ is derived as (see Appendix B)
\begin{eqnarray}\label{Linewidth}
\nonumber D&=&\frac{g^{2}\hbar\omega_{0}}{4I[(\kappa/2+\Gamma)^{2}+\Delta\omega^{2}_{BS}]}\left[R\left(\frac{3}{2}+\bar{\sigma}_{z}\right)+\frac{\Gamma}{2}{\cal{W}}_{0}\right.\\
\nonumber&&+\frac{(\kappa/2+\Gamma)\Delta\omega_{BS}}{(\kappa/2+\Gamma)^{2}+\Delta\omega^{2}_{BS}}\langle\Omega^{s}_{\textrm{eff}}(t){\cal{U}}^{s}(t)\rangle_{t}\\
&&\left.+\frac{(\kappa/2+\Gamma)^{2}-\Delta\omega^{2}_{BS}}{(\kappa/2+\Gamma)^{2}+\Delta\omega^{2}_{BS}}\langle\Omega^{s}_{\textrm{eff}}(t){\cal{V}}^{s}(t)\rangle_{t}\right].
\end{eqnarray}

Figure 3c illustrates the dependence of $D$ on $g$. When the system passes the threshold in the weak-coupling regime, $D$ is significantly suppressed, indicating the formation of maser oscillation. However, the linewidth $D$ raises as $g$ is increased in the strong-coupling regime. This is because in the limit of $(\frac{\kappa}{2}+\Gamma)\gg\Delta\omega_{BS}$, $D$ is proportional to $g^{2}\propto\eta^{2}$, meaning a small fluctuation of the atomic polarization can cause a large phase noise in the microwave field. The spectral linewidth $D$ can be even larger than both the resonator and atomic-polarization linewidths $\kappa$ and $\Gamma$, indicating the feedback (interference) effect of the resonator hardly takes effect. Surprisingly, when $g$ is further enhanced to the ultrastrong-coupling regime, $D$ starts to decrease. This is because $D\propto\eta^{-2}$ when $\Delta\omega_{BS}$ exceeds $(\frac{\kappa}{2}+\Gamma)$, i.e., the Bloch-Siegert shift $\Delta\omega_{BS}$ weakens the effect of the atomic-polarization fluctuations. We deduce that $D$ scales inversely with ${\cal{W}}_{0}$ from Fig. 3c. It should be noted that according to the Heisenberg uncertainty relation, this suppressed phase noise enhances the intensity fluctuation of the micromaser field, which is consistent with the result in~\cite{PRA:Vyas1986}. Moreover, since $D$ is inversely proportional to the intraresonator energy $I$, increasing the pumping rate $R$ always reduces $D$ (see Fig. 3d).

For the conventional masers/lasers, the quantum-limited spectral linewidth is predicted by the Schawlow-Townes formula
\begin{equation}
D_{ST}=\frac{\kappa\hbar\omega_{0}}{2I}\frac{{\cal{N}}_{u,0}}{{\cal{W}}_{0}}\left(\frac{\Gamma}{\Gamma+\kappa/2}\right)^{2},
\end{equation}
where ${\cal{N}}_{u,0}$ gives the average steady-state population in the upper masing/lasing level~\cite{PR:Schawlow1958,PRL:Kuppens1994,Book:Lax}. Applying the usual maser/laser theory to the regime of ultrastrong coupling, $D_{ST}$ can be close to or even exceed the resonator-mode frequency $\omega_{0}$ (see Fig. 3c), which is unphysical. However, if we substitute $I$, ${\cal{N}}_{u,0}$ and ${\cal{W}}_{0}$ by our numerical results without taking rotating-wave and slowly-varying-envelope approximations, the resulting $D_{ST}$ is approximately equal to $D$ (see Figs. 3c and 3d). We believe that the Schawlow-Townes linewidth $D_{ST}$ is a general form and also valid in the ultrastrong-coupling regime where all system variables should be derived without using rotating-wave and slowly-varying-envelope approximations.

\textit{Conclusion.} In summary, we have numerically studied a superconducting micromaser, where the Rydberg atoms interact with a high-$Q$ LC resonator without using rotating-wave and slowly-varying-envelope approximations. In comparison with the conventional maser/laser theory, the impact of counter-rotating terms in the atom-resonator interaction on the masing dynamics are reflected in three aspects: reducing the electric-field energy inside the superconducting resonator (see Fig. 2c), shifting the maser oscillation frequency (see Fig. 2d), and suppressing the phase fluctuation of microwave field (see Fig. 3c). Our results may be extended and tested via the quantum micromaser model based upon the continued fraction method~\cite{JPB:Stenholm1972,JPA:Swain1973-1,JPA:Swain1973-2,JPA:Swain1973-3}, which also includes the quantum effects of the electromagnetic radiation in the weak-field limit. 

Extrapolating the tremendous progress in the development of hybrid superconducting quantum circuits, our proposed system provides a promising platform for studying cavity QED in the ultrastrong-coupling regime, such as the quantum phase transition~\cite{PRL:Emary2003,PRE:Emary2003}, the squeezing of an electromagnetic field~\cite{NewJPhys:Stassi2016}, and the quantum memory~\cite{arXiv:Stassi2017}. Furthermore, it enables exploring fundamental principles in quantum optics, including the physical interpretation of the counter-rotating terms, the role of virtual photons in the atom-field interaction, and how to quantize the atom-field system entirely since the photon number is no longer a (quasi-)good quantum number when the atom-field interaction approximates (ultrastrong coupling) or even exceeds (deep strong coupling) the light frequency.

Superconducting circuits offer substantial design flexibility of resonator in the practical application. The recent experimental progress of an ensemble of Rydberg atoms interacting with a microwave cavity
~\cite{PRL:Hogan2012,arxiv:Stammeier} indicate the prospect of exploring experimentally the ultrastrong coupling regime discussed in this work.

\begin{acknowledgments}
This research is supported by the National Research Foundation Singapore under its Competitive Research Programme (CRP Award No. NRF-CRP12-2013-03) and the Centre for Quantum Technologies, Singapore.
\end{acknowledgments}

$^{\ast}$ rdumke@ntu.edu.sg

\appendix

\section{Heisenberg-Langevin Eequations}

In this Appendix, we provide the details on the derivation of equations~(\ref{rhoEq}) of motion for the macroscopic atomic variables via the Heisenberg-Langevin approach~\cite{PRA:Yu2010a,PRA:Yu2010b,PRA:Yu2011}.

From the Hamiltonian~(\ref{HamiltonianH}) and Heisenberg equation
\begin{equation}
\frac{d}{dt}O=\frac{1}{i\hbar}[O,H],
\end{equation}
for an arbitrary operator $O$, we obtain the quantum Langevin equations of motion for the operators of the $j$-th atom
\begin{subequations}
\begin{eqnarray}
\nonumber\frac{d}{dt}\sigma^{(j)}_{uu}(t)&=&-\Gamma\sigma^{(j)}_{uu}(t)-\frac{i}{2}\theta(t-t_{j})\Omega_{\textrm{eff}}(t)\\
&&\times\left[\sigma^{(j)\dag}_{-}(t)-\sigma^{(j)}_{-}(t)\right]+f^{(j)}_{\sigma_{uu}}(t),\\
\nonumber\frac{d}{dt}\sigma^{(j)}_{-}(t)&=&(-\Gamma-i\omega_{a})\sigma^{(j)}_{-}(t)+\frac{i}{2}\theta(t-t_{j})\Omega_{\textrm{eff}}(t)\\
&&\times\left[\sigma^{(j)}_{uu}(t)-\sigma^{(j)}_{ll}(t)\right]+f^{(j)}_{\sigma_{-}}(t),\\
\nonumber\frac{d}{dt}\sigma^{(j)}_{ll}(t)&=&-\Gamma\sigma^{(j)}_{ll}(t)+\frac{i}{2}\theta(t-t_{j})\Omega_{\textrm{eff}}(t)\\
&&\times\left[\sigma^{(j)\dag}_{-}(t)-\sigma^{(j)}_{-}(t)\right]+f^{(j)}_{\sigma_{ll}}(t).
\end{eqnarray}
\end{subequations}
The above Heisenberg-Langevin equations have same structure
\begin{equation}
\frac{d}{dt}x(t)=A_{x}(t)x(t)+f_{x}(t),
\end{equation}
where the Langevin forces ($\langle f_{x}^{(j)}(t)\rangle=0$) are correlated with the $\delta-$function in time, i.e.,
\begin{equation}
\langle f_{x}^{(i)}(t)f_{y}^{(j)}(t')\rangle=2d(x,y)\delta_{ij}\delta(t-t'),
\end{equation}
with the diffusion coefficient
\begin{equation}
2d(x,y)=-\langle xA_{y}\rangle-\langle A_{x}y\rangle+\frac{d}{dt}\langle xy\rangle.
\end{equation}
The non-vanishing terms are derived as
\begin{subequations}
\begin{eqnarray}
&&2d\left(\sigma_{uu},\sigma_{uu}\right)=\Gamma\langle\sigmaσ_{uu}(t)\rangle,\\
&&2d\left(\sigma_{uu},\sigma_{-}^{\dag}\right)=\Gamma\langle\sigma_{-}^{\dag}(t)\rangle,\\
&&2d\left(\sigma_{-}^{\dag},\sigma_{-}\right)=\Gamma\langle\sigma_{uu}(t)\rangle,\\
&&2d\left(\sigma_{-}^{\dag},\sigma_{ll}\right)=\Gamma\langle\sigma_{-}^{\dag}(t)\rangle,\\
&&2d\left(\sigma_{-},\sigma_{uu}\right)=\Gamma\langle\sigma_{-}(t)\rangle,\\
&&2d\left(\sigma_{-},\sigma_{-}^{\dag}\right)=\Gamma\langle\sigma_{ll}(t)\rangle,\\
&&2d\left(\sigma_{ll},\sigma_{-}\right)=\Gamma\langle\sigma_{-}(t)\rangle,\\
&&2d\left(\sigma_{ll},\sigma_{ll}\right)=\Gamma\langle\sigma_{ll}(t)\rangle.
\end{eqnarray}
\end{subequations}
Adding up the individual atomic operators, we obtain the macroscopic atomic operators
\begin{subequations}
\begin{eqnarray}
N_{u}(t)&=&\sum_{j}\theta(t-t_{j})\sigma^{(j)}_{uu}(t),\\
M^{\dag}(t)&=&\sum_{j}\theta(t-t_{j})\sigma^{(j)\dag}_{-}(t),\\
M(t)&=&\sum_{j}\theta(t-t_{j})\sigma^{(j)}_{-}(t),\\
N_{l}(t)&=&\sum_{j}\theta(t-t_{j})\sigma^{(j)}_{ll}(t),
\end{eqnarray}
\end{subequations}
for which the corresponding equations of motion are given by
\begin{subequations}
\begin{eqnarray}
\nonumber\frac{d}{dt}N_{u}(t)&=&R_{u}-\Gamma N_{u}(t)-\frac{i}{2}\Omega_{\textrm{eff}}(t)\\
&&\times\left[M^{\dag}(t)-M(t)\right]+F_{N_{u}}(t),\\
\nonumber\frac{d}{dt}M^{\dag}(t)&=&R^{\ast}_{M}+\left(-\Gamma+i\omega_{a}\right)M^{\dag}(t)-\frac{i}{2}\Omega_{\textrm{eff}}\\
&&\times\left[N_{u}(t)-N_{l}(t)\right]+F^{\dag}_{M}(t),\\
\nonumber\frac{d}{dt}M(t)&=&R_{M}+\left(-\Gamma-i\omega_{a}\right)M(t)+\frac{i}{2}\Omega_{\textrm{eff}}(t)\\
&&\times\left[N_{u}(t)-N_{l}(t)\right]+F_{M}(t),\\
\nonumber\frac{d}{dt}N_{l}(t)&=&R_{l}-\Gamma N_{l}(t)+\frac{i}{2}\Omega_{\textrm{eff}}(t)\\
&&\times\left[M^{\dag}(t)-M(t)\right]+F_{N_{l}}(t),
\end{eqnarray}
\end{subequations}
where the injection values of are given by
\begin{subequations}
\begin{eqnarray}
R_{u}&=&R\langle\sigma^{(j)}_{uu}\left(t=t_{j}\right)\rangle,\\
R_{M}&=&R\langle\sigma^{(j)\dag}_{-}\left(t=t_{j}\right)\rangle,\\
R_{l}&=&R\langle\sigma^{(j)}_{ll}\left(t=t_{j}\right)\rangle,
\end{eqnarray}
\end{subequations}
and the macroscopic Langevin forces are defined as
\begin{subequations}
\begin{eqnarray}
\nonumber F_{N_{u}}(t)&=&\sum_{j}\left[\delta(t-t_{j})\sigma^{(j)}_{uu}(t_{j})+\theta(t-t_{j})f^{(j)}_{\sigma_{uu}}(t)\right]\\
&&-R_{u},\\
\nonumber F_{M^{\dag}}(t)&=&F^{\dag}_{M}(t)\\
\nonumber&=&\sum_{j}\left[\delta(t-t_{j})\sigma^{(j)\dag}_{-}(t_{j})+\theta(t-t_{j})f^{(j)\dag}_{\sigma_{-}}(t)\right]\\
&&-R^{\ast}_{M},\\
\nonumber F_{M}(t)&=&\sum_{j}\left[\delta(t-t_{j})\sigma^{(j)}_{-}(t_{j})+\theta(t-t_{j})f^{(j)}_{\sigma_{-}}(t)\right]\\
&&-R_{M},\\
\nonumber F_{N_{l}}(t)&=&\sum_{j}\left[\delta(t-t_{j})\sigma^{(j)}_{ll}(t_{j})+\theta(t-t_{j})f^{(j)}_{\sigma_{ll}}(t)\right]\\
&&-R_{l}.
\end{eqnarray}
\end{subequations}
The above Heisenberg-Langevin equations have same structure
\begin{equation}
\frac{d}{dt}X(t)=A_{X}(t)X(t)+F_{X}(t).
\end{equation}
The macroscopic Langevin operators ($\langle F_{X}(t)\rangle=0$) are correlated with the $\delta$-function in time,
\begin{equation}
\langle F_{X}(t)F_{Y}(t')\rangle=2D(X,Y)\delta\left(t-t'\right),
\end{equation}
The non-vanishing diffusion coefficients are derived as
\begin{subequations}
\begin{eqnarray}
&&2D\left(N_{u},N_{u}\right)=R_{u}+\Gamma\langle N_{u}(t)\rangle,\\
&&2D\left(N_{u},M^{\dag}\right)=\Gamma\langle M^{\dag}(t)\rangle,\\
&&2D\left(M^{\dag},M\right)=R_{u}+\Gamma\langle N_{u}(t)\rangle,\\
&&2D\left(M^{\dag},N_{l}\right)=R^{\ast}_{M}+\Gamma\langle M^{\dag}(t)\rangle,\\
&&2D\left(M,N_{u}\right)=R_{M}+\Gamma\langle M(t)\rangle,\\
&&2D\left(M,M^{\dag}\right)=R_{l}+\Gamma\langle N_{l}(t)\rangle,\\
&&2D\left(N_{l},M\right)=R_{M}+\Gamma\langle M(t)\rangle,\\
&&2D\left(N_{l},N_{l}\right)=R_{l}+\Gamma\langle N_{l}(t)\rangle,
\end{eqnarray}
\end{subequations}
where we have chosen the Poissonian statistics,
\begin{equation}
\sum_{i\neq j}\langle\delta\left(t-t_{i}\right)\delta\left(t-t_{j}\right)\rangle_{s}=R^{2}.
\end{equation}

Now we derive the $c$-number Langevin equations, which are equivalent to the quantum Langevin equations. One should choose a particular ordering for the products of atomic operators, because the $c$-number variables commute with each other while the operators do not. Here, we choose the normal ordering of the atomic operators
\begin{equation}
\nonumber M^{\dag}(t),~N_{u}(t),~N_{l}(t),~M(t).
\end{equation}
Then we can replace the atomic operators by a set of $c$-number variables ${\cal{M}}^{\ast}(t)$, ${\cal{N}}_{u}(t)$, ${\cal{N}}_{l}(t)$, and ${\cal{M}}(t)$, respectively, which obeys the equations of motion
\begin{subequations}
\begin{eqnarray}
\nonumber\frac{d}{dt}{\cal{N}}_{u}(t)&=&R_{u}-\Gamma{\cal{N}}_{u}(t)-\frac{i}{2}\Omega_{\textrm{eff}}(t)\\
&&\times\left[{\cal{M}}^{\ast}(t)-{\cal{M}}(t)\right]+{\cal{F}}_{{\cal{N}}_{u}}(t),\\
\nonumber\frac{d}{dt}{\cal{M}}^{\ast}(t)&=&R^{\ast}_{M}+(-\Gamma+i\omega_{a}){\cal{M}}^{\ast}(t)-\frac{i}{2}\Omega_{\textrm{eff}}(t)\\
&&\times\left[{\cal{N}}_{u}(t)-{\cal{N}}_{l}(t)\right]+{\cal{F}}^{\ast}_{{\cal{M}}}(t),\\
\nonumber\frac{d}{dt}{\cal{M}}(t)&=&R_{M}+(-\Gamma-i\omega_{a}){\cal{M}}(t)+\frac{i}{2}\Omega_{\textrm{eff}}(t)\\
&&\times\left[{\cal{N}}_{u}(t)-{\cal{N}}_{l}(t)\right]+{\cal{F}}_{{\cal{M}}}(t),\\
\nonumber\frac{d}{dt}{\cal{N}}_{l}(t)&=&R_{l}-\Gamma{\cal{N}}_{l}(t)+\frac{i}{2}\Omega_{\textrm{eff}}(t)\\
&&\times\left[{\cal{M}}^{\ast}(t)-{\cal{M}}(t)\right]+{\cal{F}}_{{\cal{N}}_{l}}(t).
\end{eqnarray}
\end{subequations}
The above Langevin equations have same structure
\begin{equation}
\frac{d}{dt}{\cal{X}}(t)={\cal{A}}_{{\cal{X}}}(t){\cal{X}}(t)+{\cal{F}}_{{\cal{X}}}(t).
\end{equation}
The $c$-number Langevin forces ($\langle{\cal{F}}_{{\cal{X}}}(t)\rangle=0$) are $\delta$ correlated in time,
\begin{equation}
\langle{\cal{F}}_{{\cal{X}}}(t){\cal{F}}_{{\cal{Y}}}(t')\rangle=2{\cal{D}}({\cal{X}},{\cal{Y}})\delta(t-t'),
\end{equation}
with the diffusion coefficient
\begin{equation}
2{\cal{D}}({\cal{X}},{\cal{Y}})=2D(X,Y)+\langle XA_{Y}\rangle+\langle A_{X}Y\rangle-{\cal{A}}_{{\cal{X}}}{\cal{Y}}-{\cal{X}}{\cal{A}}_{{\cal{Y}}}.
\end{equation}
The non-vanishing diffusion coefficients are derived as
\begin{subequations}
\begin{eqnarray}
&&2{\cal{D}}\left({\cal{M}}^{\ast},{\cal{M}}^{\ast}\right)=-i\Omega_{\textrm{eff}}{\cal{M}}^{\ast}(t),\\
&&2{\cal{D}}\left({\cal{M}}^{\ast},{\cal{M}}\right)=R_{u}+\Gamma{\cal{N}}_{u}(t),\\
\nonumber&&2{\cal{D}}\left({\cal{N}}_{u},{\cal{N}}_{u}\right)=R_{u}+\Gamma{\cal{N}}_{u}(t)-\frac{i}{2}\Omega_{\textrm{eff}}(t)\\
&&~~~~~~~~~~~~~~~~~~~~\times\left[{\cal{M}}^{\ast}(t)-{\cal{M}}(t)\right],\\
&&2{\cal{D}}\left({\cal{N}}_{u},{\cal{N}}_{l}\right)=\frac{i}{2}\Omega_{\textrm{eff}}(t)\left[{\cal{M}}^{\ast}(t)-{\cal{M}}(t)\right],\\
\nonumber&&2{\cal{D}}\left({\cal{N}}_{l},{\cal{N}}_{l}\right)=R_{l}+\Gamma{\cal{N}}_{l}(t)-\frac{i}{2}\Omega_{\textrm{eff}}(t)\\
&&~~~~~~~~~~~~~~~~~~~~\times\left[{\cal{M}}^{\ast}(t)-{\cal{M}}(t)\right],\\
&&2{\cal{D}}\left({\cal{N}}_{l},{\cal{M}}\right)=R_{M}+\Gamma{\cal{M}}(t),\\
&&2{\cal{D}}\left({\cal{M}},{\cal{M}}\right)=i\Omega_{\textrm{eff}}(t){\cal{M}}(t).
\end{eqnarray}
\end{subequations}

We further define the $c$-number macroscopic variables
\begin{subequations}
\begin{eqnarray}
{\cal{U}}(t)&=&{\cal{M}}^{\ast}(t)+{\cal{M}}(t),\\
{\cal{V}}(t)&=&-i\left[{\cal{M}}^{\ast}(t)-{\cal{M}}(t)\right],\\
{\cal{W}}(t)&=&{\cal{N}}_{u}(t)-{\cal{N}}_{l}(t),
\end{eqnarray}
\end{subequations}
the injection parameters
\begin{subequations}
\begin{eqnarray}
R_{{\cal{U}}}&=&R^{\ast}_{M}+R_{M},\\
R_{{\cal{V}}}&=&-i\left(R^{\ast}_{M}-R_{M}\right),\\
R_{{\cal{W}}}&=&R_{u}-R_{l},
\end{eqnarray}
\end{subequations}
and the $c$-number macroscopic Langevin forces
\begin{subequations}
\begin{eqnarray}
{\cal{F}}_{{\cal{U}}}(t)&=&{\cal{F}}^{\ast}_{{\cal{M}}}(t)+{\cal{F}}_{{\cal{M}}}(t),\\
\nonumber{\cal{F}}_{{\cal{V}}}(t)&=&-i\left[{\cal{F}}^{\ast}_{{\cal{M}}}(t)-{\cal{F}}_{{\cal{M}}}(t)\right]\\
&=&-i\left[{\cal{F}}_{{\cal{M}}^{\ast}}(t)-{\cal{F}}_{{\cal{M}}}(t)\right],\\
{\cal{F}}_{{\cal{W}}}(t)&=&{\cal{F}}_{{\cal{N}}_{u}}(t)-{\cal{F}}_{{\cal{N}}_{l}}(t),
\end{eqnarray}
\end{subequations}
and obtain the $c$-number Langevin equations, which correspond to Eq.~(\ref{rhoEq}),
\begin{subequations}\label{cnumberequations}
\begin{eqnarray}
\frac{d}{dt}{\cal{U}}(t)&=&R_{{\cal{U}}}-\Gamma{\cal{U}}(t)-\omega_{a}{\cal{V}}(t)+{\cal{F}}_{{\cal{U}}}(t),\\
\nonumber\frac{d}{dt}{\cal{V}}(t)&=&-\Gamma{\cal{V}}(t)+\omega_{a}{\cal{U}}(t)-\Omega_{\textrm{eff}}(t){\cal{W}}(t)\\
&&+{\cal{F}}_{{\cal{V}}}(t),\\
\frac{d}{dt}{\cal{W}}(t)&=&R_{{\cal{W}}}-\Gamma{\cal{W}}(t)+\Omega_{\textrm{eff}}(t){\cal{V}}(t)+{\cal{F}}_{{\cal{W}}}(t),~~~~~
\end{eqnarray}
\end{subequations}
with $R_{{\cal{U}}}=R\bar{\sigma}_{x}$, $R_{{\cal{V}}}=0$, and $R_{{\cal{W}}}=R\bar{\sigma}_{z}$.

For an arbitrary set of initial states, the Rydberg micromaser system arrives at a steady state after a long enough time $t_{0}$. These steady-state solutions oscillate at the common frequency $\omega_{c}$ and, therefore, can be expanded into the Fourier series, i.e.,
\begin{subequations}
\begin{eqnarray}
{\cal{E}}^{s}(t)&=&\frac{1}{2}{\cal{E}}_{0}e^{i\omega_{c}t}+\frac{1}{2}{\cal{E}}_{0}e^{-i\omega_{c}t},\\
{\cal{U}}^{s}(t)&=&\sum_{n}u^{s}_{n}e^{in\omega_{c}t},\\
{\cal{V}}^{s}(t)&=&\sum_{n}v^{s}_{n}e^{in\omega_{c}t},\\
{\cal{W}}^{s}(t)&=&\sum_{n}w^{s}_{n}e^{in\omega_{c}t}.
\end{eqnarray}
\end{subequations}
We use the superscript '$s$' to denote the steady state. ${\cal{E}}_{0}$ is the amplitude of the electric-field oscillation. The steady-state effective Rabi frequency is then given by
\begin{equation}
\Omega_{\textrm{eff}}^{s}(t)=\sum_{n=-1,0,1}q^{s}_{n}e^{in\omega_{c}t},
\end{equation}
with $q^{s}_{0}=\Omega$, $q^{s}_{1}=-\mathpzc{d}_{0}{\cal{E}}_{0}/\hbar$, $q^{s}_{-1}=-\mathpzc{d}_{0}{\cal{E}}_{0}/\hbar$.

In the steady state, we consider the average values of different atomic variables, i.e.,
\begin{subequations}
\begin{eqnarray}
\nonumber{\cal{W}}_{0}&=&\langle{\cal{W}}^{s}(t)\rangle_{t}\\
&=&\lim_{T\to\infty}\frac{\epsilon_{0}V}{T}\int_{t_{0}}^{t_{0}+T}{\cal{W}}^{s}(t)dt,
\end{eqnarray}
\end{subequations}
${\cal{N}}_{u,0}=\langle{\cal{N}}^{s}_{u}(t)\rangle_{t}$, ${\cal{N}}_{l,0}=\langle{\cal{N}}^{s}_{l}(t)\rangle_{t}$, and ${\cal{W}}_{0}={\cal{N}}_{u,0}-{\cal{N}}_{l,0}$. The conservation of energy and conservation of atomic number lead to the equations
\begin{subequations}
\begin{eqnarray}
\hbar\omega_{a}\left(R_{u}-\Gamma{\cal{N}}_{u,0}\right)&=&\kappa I,\\
\Gamma\left({\cal{N}}_{u,0}+{\cal{N}}_{l,0}\right)&=&R,
\end{eqnarray}
\end{subequations}
from which one can easily obtain Eq.~(\ref{relation1}).

\section{Spectral linewidth of Rydberg micromaser}

To investigate the spectral linewidth of the microwave, we write the electric field and different atomic variables as a sum of the steady-state solution and a small fluctuation,
\begin{subequations}
\begin{eqnarray}
&&{\cal{E}}(t)={\cal{E}}^{s}(t)+\delta{\cal{E}}(t),\\
&&{\cal{U}}(t)={\cal{U}}^{s}(t)+\delta{\cal{U}}(t),\\
&&{\cal{V}}(t)={\cal{V}}^{s}(t)+\delta{\cal{V}}(t),\\
&&{\cal{W}}(t)={\cal{W}}^{s}(t)+\delta{\cal{W}}(t).
\end{eqnarray}
\end{subequations}
The effective Rabi frequency is written as
\begin{equation}
\Omega_{\textrm{eff}}(t)=\Omega^{s}_{\textrm{eff}}(t)-\frac{2\mathpzc{d}_{0}}{\hbar}\delta{\cal{E}}(t).
\end{equation}
It is easy to derive the following equations of motion for the small fluctuations
\begin{subequations}
\begin{eqnarray}
\frac{d}{dt}\delta{\cal{U}}(t)&=&-\Gamma\delta{\cal{U}}(t)-\omega_{a}\delta{\cal{V}}(t)+{\cal{F}}_{{\cal{U}}}(t),\\
\nonumber\frac{d}{dt}\delta{\cal{V}}(t)&=&-\Gamma\delta{\cal{V}}(t)+\omega_{a}\delta{\cal{U}}(t)-\Omega^{s}_{\textrm{eff}}(t)\delta{\cal{W}}(t)\\
&&-\frac{2\mathpzc{d}_{0}}{\hbar}{\cal{W}}^{s}(t)\delta{\cal{E}}(t)+{\cal{F}}_{{\cal{V}}}(t),\\
\nonumber\frac{d}{dt}\delta{\cal{W}}(t)&=&-\Gamma\delta{\cal{W}}(t)+\Omega^{s}_{\textrm{eff}}(t)\delta{\cal{V}}(t)\\
&&-\frac{2\mathpzc{d}_{0}}{\hbar}{\cal{V}}^{s}(t)\delta{\cal{E}}(t)+{\cal{F}}_{{\cal{W}}}(t).
\end{eqnarray}
\end{subequations}
We then apply the Fourier transform to convert above differential equations into a set of algebraic equations
\begin{subequations}
\begin{eqnarray}
\delta{\cal{V}}(\omega)&=&\frac{\Gamma+i\omega}{\omega_{a}}\frac{\epsilon_{0}V_{\textrm{eff}}}{\mathpzc{d}_{0}}\Pi(\omega)\delta{\cal{E}}(\omega)+\frac{{\cal{F}}_{{\cal{U}}}(\omega)}{\omega_{a}},\\
\nonumber\delta{\cal{V}}(\omega)&=&-\frac{\omega_{a}}{\Gamma+i\omega}\frac{\epsilon_{0}V_{\textrm{eff}}}{\mathpzc{d}_{0}}\Pi(\omega)\delta{\cal{E}}(\omega)\\
\nonumber&&-\sum_{n}\frac{q^{s}_{n}}{\Gamma+i\omega}\delta{\cal{W}}(\omega-n\omega_{c})+\frac{{\cal{F}}_{{\cal{V}}}(\omega)}{\Gamma+i\omega}\\
&&-\frac{2\mathpzc{d}_{0}}{\hbar}\sum_{n}\frac{w^{s}_{n}}{\Gamma+i\omega}\delta{\cal{E}}(\omega-n\omega_{c}),\\
\nonumber\delta{\cal{W}}(\omega)&=&\sum_{n}\frac{q^{s}_{n}}{\Gamma+i\omega}\delta{\cal{V}}(\omega-n\omega_{c})+\frac{{\cal{F}}_{{\cal{W}}}(\omega)}{\Gamma+i\omega}\\
&&-\frac{2\mathpzc{d}_{0}}{\hbar}\sum_{n}\frac{w^{s}_{n}}{\Gamma+i\omega}\delta{\cal{E}}(\omega-n\omega_{c}),
\end{eqnarray}
\end{subequations}
where we have used the equation
\begin{equation}
\delta{\cal{U}}(\omega)=-\frac{\epsilon_{0}V_{\textrm{eff}}}{\mathpzc{d}_{0}}\Pi(\omega)\delta{\cal{E}}(\omega),
\end{equation}
which is derived from the Fourier transform of the wave equation~(\ref{ElectricFieldEq}) for the electric field, with
\begin{equation}
\Pi(\omega)=\frac{(i\omega)^{2}+\kappa(i\omega)+\omega_{0}^{2}}{(i\omega)^{2}}.
\end{equation}

For a maser/laser operates above the threshold, the spectral linewidth is main from the microwave phase fluctuations, Eq.~(\ref{phasefluctuation}). We focus the region of $\omega$ close to $\omega_{c}$, and $\delta\varphi(\omega)$ can be expressed as
\begin{subequations}
\begin{eqnarray}
\nonumber\delta\varphi(\omega)&\approx&\frac{1}{\omega-\omega_{c}}\frac{\mathpzc{d}_{0}}{2\epsilon_{0}V_{\textrm{eff}}{\cal{E}}_{0}}\\
&&\times\frac{\Delta\omega_{BS}{\cal{F}}_{{\cal{U}}}(\omega)+(\kappa/2+\Gamma){\cal{F}}_{{\cal{V}}}(\omega)}{(\kappa/2+\Gamma)^{2}+\Delta\omega_{BS}^{2}},
\end{eqnarray}
\end{subequations}
where we has used the approximations
\begin{subequations}
\begin{eqnarray}
&&\Pi(\omega)\approx(2/\omega_{0})(\omega-\omega_{0}-i\kappa/2),\\
&&\omega_{a}^{2}+(\Gamma+i\omega)^{2}\approx-2\omega_{c}\left[(\omega-\omega_{a})-i\Gamma\right].
\end{eqnarray}
\end{subequations}
Based on the correlation properties of ${\cal{F}}^{\ast}_{{\cal{M}}}(\omega)$ and ${\cal{F}}_{{\cal{M}}}(\omega)$, the phase noise $\delta\varphi(\omega)$ is $\delta$-correlated in frequency, i.e., Eq.~(\ref{deltacorrelated}), and the spectral linewidth is given by Eq.~(\ref{Linewidth}).


\begin{thebibliography}{10}

\bibitem{PRA:Ciuti2005} C. Ciuti, G. Bastard, and I. Carusotto, Phys. Rev. B {\bf 72}, 115303 (2005).

\bibitem{PRA:Ciuti2006} C. Ciuti and I. Carusotto, Phys. Rev. A {\bf 74}, 033811 (2006).

\bibitem{AnnPhys:Devoret2007} M. Devoret, S. Girvin, and R. Schoelkopf, Ann. Phys. (Leipzig) {\bf 16}, 767 (2007).

\bibitem{NatPhys:Niemczyk2010} T. Niemczyk, F. Deppe, H. Huebl, E. P. Menzel, F. Hocke, M. J. Schwarz, J. J. Garcia-Ripoll, D. Zueco, T. H\"ummer, E. Solano, A. Marx, and R. Gross, Nat. Phys. {\bf 6}, 772 (2010).

\bibitem{EPL:Haroch2008} C. Roux, A. Emmert, A. Lupascu, T. Nirrengarten, G. Nogues, M. Brune, J.-M. Raimond, and S. Haroche, Eur. Phys. Lett. {\bf 81}, 56004 (2008).

\bibitem{PRL:Shimizu2009} F. Shimizu, C. Hufnagel, and T. Mukai, Phys. Rev. Lett. {\bf 103}, 253002 (2009).

\bibitem{PRA:Siercke2012} M. Siercke, K. S. Chan, B. Zhang, M. Beian, M. J. Lim, and R. Dumke, Phys. Rev. A {\bf 85}, 041403 (2012).

\bibitem{PRA:Patton2013} K. R. Patton and U. R. Fischer, Phys. Rev. A {\bf 87}, 052303 (2013).

\bibitem{NatComm:Bernon2013} S. Bernon, H. Hattermann, D. Bothner, M. Knufinke, P. Weiss, F. Jessen, D. Cano, M. Kemmler, R. Kleiner, D. Koelle, and J. Fort\'agh, Nat. Commun. {\bf 4}, 2380 (2013).


\bibitem{PRA:Yu2016} D. Yu, M. M. Valado, C. Hufnagel, L. C. Kwek, L. Amico, and R. Dumke, Phys. Rev. A {\bf 93}, 042329 (2016).

\bibitem{PR:Lamb1964} W. E. Lamb, Jr., Phys. Rev. {\bf 134}, A1429 (1964).

\bibitem{PRL:Scully1966} M. Scully and W. E. Lamb, Jr., Phys. Rev. Lett. {\bf 16}, 853 (1966).

\bibitem{PR:Stenholm1969} S. Stenholm and W. E. Lamb, Jr., Phys. Rev. {\bf 181}, 618 (1969).

\bibitem{PRA:Benkert1990} C. Benkert, M. O. Scully, J. Bergou, L. Davidovich, M. Hillery, and M. Orszag, Phys. Rev. A {\bf 41}, 2756 (1990).

\bibitem{PRA:Kolobov1993} M. I. Kolobov, L. Davidovich, E. Giacobino, and C. Fabre, Phys. Rev. A {\bf 47}, 1431 (1993).

\bibitem{RMP:Davidovich1996} L. Davidovich, Rev. Mod. Phys. {\bf 68}, 127 (1996).

\bibitem{PRA:Filipowicz1986} P. Filipowicz, J. Javanainen, and P. Meystre, Phys. Rev. A {\bf 34}, 3077 (1986).

\bibitem{PRL:Rempe1987} Gerhard Rempe, Herbert Walther, and Norbert Klein, Phys. Rev. Lett. {\bf 58}, 353 (1987).


\bibitem{PRL:An1994} K. An, J. J. Childs, R. R. Dasari, and M. S. Feld, Phys. Rev. Lett. {\bf 73}, 3375 (1994).


\bibitem{Nature:McKeever2003} J. McKeever, A. Boca, A. D. Boozer, J. R. Buck, and H. J. Kimble, Nature (London) {\bf 425}, 268 (2003).

\bibitem{JOptSocAmB:Clemens2004} J. P. Clemens, P. R. Rice, and L. M. Pedrotti, J. Opt. Soc. Am. B. {\bf 21}, 2025 (2004).

\bibitem{PRA:FangYen2006} C. Fang-Yen, C. C. Yu, S. Ha, W. Choi, K. An, R. R. Dasari, and M. S. Feld, Phys. Rev. A {\bf 73}, 041802 (2006).

\bibitem{JOSAb:Yu2016} D. Yu, J. Opt. Soc. Am. B {\bf 33}, 797 (2016).

\bibitem{PhysScr:Larson2007} J. Larson, Phys. Scr. {\bf 76}, 146 (2007).

\bibitem{PhysicaA:Ford1997} G. W. Ford and R. F. O’Connell, Physica A {\bf 243}, 377 (1997).

\bibitem{JETP:Belobrov1976} P. I. Belobrov, G. M. Zaslavskii, and G. Kh. Tartakovskii, Sov. Phys. JETP {\bf 44}, 945 (1976)

\bibitem{PRL:Milonni1983} P. W. Milonni, J. R. Ackerhalt, and H. W. Galbraith, Phys. Rev. Lett. {\bf 50}, 966 (1983).

\bibitem{PRA:Peng1992} J.-S. Peng and G.-X. Li, Phys. Rev. A {\bf 45}, 3289 (1992).

\bibitem{PRL:Emary2003} C. Emary and T. Brandes, Phys. Rev. Lett. {\bf 90}, 044101 (2003).

\bibitem{PRE:Emary2003} C. Emary and T. Brandes, Phys. Rev. E {\bf 67}, 066203 (2003).

\bibitem{EurophysLett:Saito2006} K. Saito, M. Wubs, S. Kohler, P. H\"anggi, and Y. Kayanuma, Europhys. Lett. {\bf 76}, 22 (2006).

\bibitem{PRB:Saito2007} K. Saito, M. Wubs, S. Kohler, Y. Kayanuma, and P. H\"anggi, Phys. Rev. B {\bf 75}, 214308 (2007).

\bibitem{PRA:Jing2009} J. Jing, Z.-G. L\"u, and Z. Ficek, Phys. Rev. A {\bf 79}, 044305 (2009).

\bibitem{PhysScr:Ficek2010} Z. Ficek, J. Jing, and Z. G. L\"u, Phys. Scr. {\bf T140}, 014005 (2010).

\bibitem{PRA:Vyas1986} R. Vyas and S. Singh, Phys. Rev. A {\bf 33}, 375 (1986).

\bibitem{OptCommun:Zela1997} F. D. Zela, E. Solano, and A. Gago, Opt. Commun. {\bf 142}, 106 (1997).

\bibitem{JModOpt:Liu1998} Z. Liu, L. Zeng, and S. Zhu, J. Mod. Opt. {\bf 45}, 945 (1997).

\bibitem{Carter:Zela2001} F. De Zela, in \textit{Modern Challenges in Quantum Optics}, edited by M. Orszag and J. C. Retamal (Berlin, Springer, 2001). 

\bibitem{arXiv:Sarabi2017} The inductor can be designed in a single layer spiral coil structure as in the article: B. Sarabi, P. Huang, N. M. Zimmerman, arXiv:1702.02210 (2017).







\bibitem{PRA:Blais2004} A. Blais, R.-S. Huang, A. Wallraff, S. M. Girvin, and R. J. Schoelkopf, Phys. Rev. A {\bf 69}, 062320 (2004).

\bibitem{Book:Scully} M. O. Scully and M. S. Zubairy, \emph{Quantum Optics} (Cambridge University Press, Cambridge, England, 1997).

\bibitem{PRA:Zimmerman1979} M. L. Zimmerman, M. G. Littman, M. M. Kash, and D. Kleppner, Phys. Rev. A {\bf 20}, 2251 (1979).

\bibitem{PRA:Marinescu1994} M. Marinescu, H. R. Sadeghpour, and A. Dalgarno, Phys. Rev. A {\bf 49}, 982 (1994).

\bibitem{PRA:Yu2016-2} D. Yu, A. Landra, M. M. Valado, C. Hufnagel, L. C. Kwek, L. Amico, and R. Dumke, Phys. Rev. A {\bf 94}, 062301 (2016).

\bibitem{PRL:Vliegen2004} E. Vliegen, H. J. W\"orner, T. P. Softley, and F. Merkt, Phys. Rev. Lett. {\bf 92}, 033005 (2004).



\bibitem{JPB:Branden} D. B. Branden, T. Juhasz, T. Mahlokozera, C. Vesa, R. O. Wilson, M. Zheng, A. Kortyna, and D. A. Tate, J. Phys. B: At., Mol. Opt. Phys. {\bf 43}, 015002 (2010).

\bibitem{PRA:HermannAvigliano2014} C. Hermann-Avigliano, R. C. Teixeira, T. L. Nguyen, T. Cantat-Moltrecht, G. Nogues, I. Dotsenko, S. Gleyzes, J. M. Raimond, S. Haroche, and M. Brune, Phys. Rev. A {\bf 90}, 040502 (2014).

\bibitem{EPJST:Kohlhoff2016} M. W. Kohlhoff, Eur. Phys. J. Spec. Top. {\bf 225}, 3061 (2016).

\bibitem{PRA:Crosse2010} J. A. Crosse, S. \AA{}. Ellingsen, Kate Clements, Stefan Y. Buhmann, and Stefan Scheel, Phys. Rev. A {\bf 82}, 010901 (2010).

\bibitem{JPB:Fabre1983} C. Fabre, M. Gross, J. M. Raimond, and S. Haroche, J. Phys. B: At. Mol. Phys. {\bf 16}, L671 (1983).

\bibitem{PRA:Yu2010a} D. Yu and J. Chen, Phys. Rev. A {\bf 81}, 023818 (2010).

\bibitem{PRA:Yu2010b} D. Yu and J. Chen, Phys. Rev. A {\bf 81}, 053809 (2010).

\bibitem{PRA:Yu2011} D. Yu and J. Chen, Phys. Rev. A {\bf 83}, 063846 (2011).

\bibitem{PRA:Yu2008} D. Yu and J. Chen, Phys. Rev. A {\bf 78}, 013846 (2008).

\bibitem{PhysScr:Kimble1998} H. J. Kimble, Phys. Scr. {\bf T76}, 127 (1998).

\bibitem{Book:Ohtsubo} J. Ohtsubo, \textit{Semiconductor Lasers: Stability, Instability and Chaos} (Springer, 2013). 

\bibitem{PR:Bloch1940} F. Bloch and A. Siegert, Phys. Rev. {\bf 57}, 522 (1940).

\bibitem{PRB:Shirley1965} J. H. Shirley, Phys. Rev. {\bf 138}, B979 (1965).

\bibitem{PRL:FornDiaz2010} P. Forn-D\'{\i}az, J. Lisenfeld, D. Marcos, J. J. Garc\'{\i}a-Ripoll, E. Solano, C. J. P. M. Harmans, and J. E. Mooij, Phys. Rev. Lett. {\bf 105}, 237001 (2010).

\bibitem{PRL:Tuorila2010} J. Tuorila, M. Silveri, M. Sillanp\"a\"a, E. Thuneberg, Y. Makhlin, and P. Hakonen, Phys. Rev. Lett. {\bf 105}, 257003 (2010).

\bibitem{Book:Lasers-Siegman} A. E. Siegman, \textit{Lasers} (University Science Books, Sausalito, CA, 1986).

\bibitem{PR:Schawlow1958} A. L. Schawlow and C. H. Townes, Phys. Rev. {\bf 112}, 1940 (1958).

\bibitem{PRL:Kuppens1994} S. J. M. Kuppens, M. P. van Exter, and J. P. Woerdman, Phys. Rev. Lett. {\bf 72}, 3815 (1994).

\bibitem{Book:Lax} M. Lax, in \textit{Physics of Quantum Electronics}, edited by P. L. Kelley, B. Lax, and P. E. Tannenwald (McGraw-Hill, New York, 1966). 

\bibitem{JPB:Stenholm1972} S. Stenholm, J. Phys. B: At. Mol. Phys. {\bf 5}, 878 (1972).

\bibitem{JPA:Swain1973-1} S. Swain, J. Phys. A: Math. Nucl. Gen. {\bf 6}, 192 (1973).

\bibitem{JPA:Swain1973-2} S. Swain, J. Phys. A: Math. Nucl. Gen. {\bf 6}, 1919 (1973).

\bibitem{JPA:Swain1973-3} S. Swain, J. Phys. A: Math. Gen. {\bf 8}, 1277 (1975).

\bibitem{NewJPhys:Stassi2016} R. Stassi, S. Savasta, L. Garziano, B. Spagnolo, and F. Nori, New J. Phys. {\bf 18}, 123005 (2016).

\bibitem{arXiv:Stassi2017} R. Stassi and F. Nori, arXiv:1703.08951 (2017).


\bibitem{PRL:Hogan2012} S. D. Hogan, J. A. Agner, F. Merkt, T. Thiele, S. Filipp, and A. Wallraff, Phys. Rev. Lett. {\bf 108}, 063004 (2012).

\bibitem{arxiv:Stammeier} M. Stammeier, S. Garcia, T. Thiele, J. Deiglmayr, J. A. Agner, H. Schmutz, F. Merkt, and A. Wallraff, arXiv:1701.01426 (2017).

\end{thebibliography}
\end{document}